\documentclass[11pt]{article}
\usepackage{setstack}

\usepackage{graphicx}
\usepackage{epstopdf}

\usepackage{amssymb}
\usepackage{mathptmx}

\usepackage{subfigure}

\newcommand{\comments}[1]{}

\newcommand{\be}{\begin{equation}}
\newcommand{\ee}{\end{equation}}

\begin{document}

\title{\textbf{Dimer and fermionic formulations \\ of a class of colouring problems}}

\author{J. O. Fj{\ae}restad
\medskip \\ 
The University of Queensland, School of Mathematics and Physics,\\
Brisbane, QLD 4072, Australia}
\smallskip 
 
\maketitle 
 
\begin{abstract} 
\noindent

We show that the number $Z$ of $q$-edge-colourings of a simple regular graph of degree $q$ is deducible from functions describing dimers on the
same graph, viz. the dimer generating function or equivalently the set of connected dimer correlation functions. Using this relationship to the dimer problem, we derive fermionic representations for $Z$ in terms of Grassmann integrals with quartic actions. Expressions are given for planar graphs and for nonplanar graphs embeddable (without edge crossings) on a torus. We discuss exact numerical evaluations of the Grassmann integrals using an algorithm by Creutz, and present an application to the 4-edge-colouring problem on toroidal square lattices, comparing the results to numerical transfer matrix calculations and a previous Bethe ansatz study. We also show that for the square, honeycomb, 3-12, and one-dimensional lattice, known exact results for the asymptotic scaling of $Z$ with the number of vertices can be expressed in a unified way as different values of one and the same function.

\end{abstract}

\section{Introduction}
\label{intro}

Some problems in classical statistical mechanics have been found to have alternative formulations involving fermionic (i.e. anticommuting) degrees of freedom.  
The most well-known examples are the Ising and dimer models \cite{reviews} (which are related, as the former model can be mapped to the latter \cite{fisher-66}).
For planar graphs the generating function for the dimer model can be written as a Pfaffian \cite{kast-63}, which can be regarded as a partition function
for a system of \textit{noninteracting} fermionic degrees of freedom \cite{samuel-1}. The Pfaffian method can be generalized to nonplanar graphs of genus $g$, for which the dimer generating function can
be written as a linear combination of $4^g$ Pfaffians, as first posited in Ref. \cite{kast-61}, where this solution was explicitly demonstrated for the case of a square lattice embedded on a torus ($g=1$); more
general and rigorous discussions and proofs have later been given \cite{dolbilin,regge,galluccio,tesler,cimasoni}. The Pfaffian method allows for a straightforward solution of the Ising and dimer models for 
two-dimensional lattices without crossing edges, as then the partition function involves only one or at most a few Pfaffians, and explicit expressions for these can be easily found using Fourier transformation
provided the parameters describing the spin-spin interactions/dimer weights have some translational invariance.  

More recently, some other classical spin models \cite{clusel}, as well as generating functions for certain types of forests and trees (some of which are related to the $q\to 0$ limit of antiferromagnetic Potts models) \cite{caracciolo-04}, have also been shown to have fermionic formulations. Notably, for these problems the fermions are \textit{interacting}, so that even if one limits consideration to planar graphs, finding explicit exact solutions is not expected to be possible in general. 

For all problems mentioned above, the fermionic formulations are most naturally given in terms of integrals over Grassmann variables \cite{samuel-1} living on the vertices of some graph. A Pfaffian (whose square is
a determinant) can be expressed as a Gaussian Grassmann integral, i.e. its ``action" is a quadratic form in the Grassmann variables, corresponding to noninteracting fermionic degrees of freedom.  
In contrast, in the interacting case the action is non-Gaussian (typically quartic).  

In this paper we find fermionic formulations of another class of problems: the enumeration of $q$-edge-colourings of simple regular graphs of degree $q$. 
If each edge of a graph is coloured with one out of $q$ possible colours, such that no edges connected to the same vertex have the same colour, the result is said to be a $q$-edge-colouring of the graph. 
A simple regular graph of degree $q$ (from now on referred to as a graph of degree $q$ for short) is a simple graph for which all vertices are connected to exactly $q$ edges. We also note here that the 
$q$-edge-colouring problem on a graph can be reformulated as a $q$-vertex-colouring problem (i.e. a zero-temperature $q$-state antiferromagnetic Potts model) on a closely related graph (see e.g. \cite{jof10}). 

We first show that the number of $q$-edge-colourings $Z$ of a graph of degree $q$ can be obtained from the dimer generating function ${\cal Z}$ for the same graph by successively differentiating ${\cal Z}^q$ with respect to the dimer weight on each edge. This expression for $Z$ can alternatively be rewritten in terms of connected dimer correlation functions. The Grassmann integral representation for $Z$ is obtained by invoking the Pfaffian solution for ${\cal Z}$ expressed in terms of Grassmann integrals. We consider planar graphs (including graphs that can be embedded on a cylinder) as well as (nonplanar) graphs that can be embedded on a torus. The Grassmann integral expressions obtained for $Z$ for these two types of graphs differ due to the difference in the forms of the Pfaffian solution for ${\cal Z}$, the origin of which is the different topology of the embedded graphs, as mentioned above.

Our final Grassmann integral expressions for $Z$ (Eqs. (\ref{Z-main-2}) and (\ref{Z-t-main-2})) involve $q$ Grassmann variables (one for each colour) on each vertex and 2 Grassmann variables on each edge. Each term in the action couples the Grassmann variables on an edge to the Grassmann variables with a given colour index on the two vertices connected to the edge. Thus the action is spatially local and quartic. The action does not couple different colours directly; instead there is an indirect coupling via the edge Grassmann variables.

The Grassmann integral formulation could serve as a starting point for further work based on analytical or numerical approaches. As an example of the latter, we discuss numerically exact evaluations of the Grassmann integrals
using an algorithm by Creutz \cite{creutz}. Applying this approach to the enumeration of $4$-edge-colourings on finite-size square lattices embedded on a torus, we verify that the results are in agreement with numerical transfer matrix 
calculations and also consistent with an earlier Bethe ansatz prediction for the thermodynamic limit \cite{dc-nien-04}. Another
possible approach (not pursued here) might be to make use of Hubbard-Stratonovich or related transformations (see e.g. \cite{lee-lee}) to map the fermionic problem into a bosonic one (i.e. one 
involving integrals over ordinary $c$-numbers), which could then be analyzed using various methods.

Exact results for the asymptotic exponential growth of $Z$ with the number of vertices $N$ have previously been derived for the honeycomb \cite{baxter-70}, square \cite{dc-nien-04}, 3-12 \cite{jof10}, and 4-8 \cite{jof10} lattices. 
In an Appendix we show that the results for the first three of these, and for the one-dimensional chain, can be written in a unified way as different values of one and the same function. 
This suggests that a unified derivation of this asymptotic growth might be possible at least for this collection of lattices. We hope that the Grassmann integral formulations developed in this paper might be useful for shedding further light on this issue.

This paper is organized as follows. Sec. \ref{dimer} discusses the dimer formulations, with a technical derivation relegated to Appendix \ref{proof}. The fermionic (Grassmann integral) formulations are discussed in Sec. \ref{fermionic}. 
The numerical evaluation of the number of 4-edge-colourings on the toroidal square lattice is presented in Sec. \ref{num}, with some remarks on the computer implementation in Appendix \ref{impl}. Concluding remarks are given in Sec. \ref{remarks}. 
Appendix \ref{unifying} presents the unifying formula for previous asymptotic results for $Z$ for some lattices.

\section{Dimer formulations}
\label{dimer}

\subsection{Formulation in terms of the dimer generating function}
\label{dimerZ}

Consider an undirected graph $G=(V,E)$ where $V$ is the set of vertices and $E$ the set of edges. We assume that the graph is simple (i.e. each edge connects two different vertices and any two vertices are connected by at most one edge)
and that all vertices have the same degree $q$ (i.e. each vertex has exactly $q$ edges connected to it). Thus $G$ is a simple regular graph of degree $q$; for short, we will refer to $G$ as a graph of degree $q$.

Labeling the edges as $\alpha=1,\ldots,|E|$ where $|E|$ is the total number of edges, a dimer covering $n$ of $G$ can be represented as $n=(n_1,n_2,\ldots,n_{|E|})$ where $n_{\alpha}$ ($=0$ or $1$) is the
number of dimers on edge $\alpha$, subject to the constraint that exactly one of the edges connected to each vertex hosts a dimer. Letting $w_{\alpha}$ denote the weight of a dimer on edge $\alpha$, the dimer partition function 
(generating function) ${\cal Z}$ is
\be
{\cal Z} = \sum_{n}\prod_{\alpha}w_{\alpha}^{n_{\alpha}}.
\ee
Evaluating ${\cal Z}$ with all dimer weights set equal to 1 gives the total number of dimer coverings.

Next, we introduce coloured dimers and define a $c$-coloured dimer covering to consist of dimers that all have the same colour $c$, which may take one out of $q$ values: $c=1,2,\ldots, q$. Furthermore, we define a $q$-dimer-covering of $G$ to be a composite structure of $q$ dimer coverings, one of each colour. Because all vertices have degree $q$, any $q$-edge-colouring of $G$ is also a $q$-dimer-covering, but unlike generic $q$-dimer-coverings it satisfies the additional ``colouring constraint" that no edge should have more than one dimer, or, equivalently, no edge should lack a dimer. Thus the set of $q$-edge-colourings is a subset of the set of $q$-dimer-coverings. We therefore want to extract the former from the latter. To this end, consider the generating function for $q$-dimer-coverings, which is given by
\be
{\cal Z}^q = \sum_{n^{(1)},\ldots, n^{(q)}}\prod_{\alpha}w_{\alpha}^{\sum_{c=1}^q n_{\alpha}^{(c)}}
\label{Zq}
\ee
where $n^{(c)}$ denotes a $c$-coloured dimer covering. Each term in the sum corresponds to a $q$-dimer-covering. A term that also corresponds to a $q$-edge-colouring will contain exactly one factor of $w_{\alpha}$ for each edge $\alpha$. Conversely, for a term that does not correspond to a $q$-edge-colouring, at least one edge $\alpha$ will contain more than one dimer and thus comes with a factor $w_{\alpha}^p$ with $p\geq 2$, and, equivalently, at least one edge $\beta$ will not have a dimer and thus comes with a factor $w_{\beta}^0=1$. The latter property implies that if we successively differentiate a term in (\ref{Zq}) with respect to \textit{each} dimer weight $w_{\gamma}$, the final result will be 1 for any term representing a $q$-edge-colouring and 0 for all other terms. Therefore the number of $q$-edge-colourings $Z$ is given by
\be
Z = \left(\prod_{\alpha}\frac{\partial }{\partial w_{\alpha}}\right){\cal Z}^q.
\label{Zqcol}
\ee
We have thus shown that the number of $q$-edge-colourings $Z$ can be obtained from the generating function ${\cal Z}$ for the dimer problem with edge-dependent dimer weights on the same graph.

\subsection{Formulation in terms of connected dimer correlation functions}
\label{dimercorr}

In this subsection we will see that the number of $q$-edge-colourings $Z$ can be expressed in terms of connected correlation functions for the dimer problem. (Readers who are mainly
interested in the Grassmann integral formulations of $Z$ may skip this subsection, as later sections do not depend on it.)

The $m$-point dimer correlation function for the $m$ edges $\alpha_1,\ldots,\alpha_m$ (where $m$ can take values from $1$ to $|E|$) is given by
\begin{eqnarray}
{\cal G}^{(m)}(\alpha_1,\ldots,\alpha_m) & \equiv &  \langle n_{\alpha_1}n_{\alpha_2}\cdots n_{\alpha_m}\rangle 
= \frac{1}{\cal Z}\sum_{n}n_{\alpha_1}\cdots n_{\alpha_m}\;\prod_{\alpha}w_{\alpha}^{n_{\alpha}} \nonumber \\ &=& \frac{1}{\cal Z}\left(\prod_{k=1}^m \frac{\partial}{\partial \log w_{\alpha_k}}\right){\cal Z}.
\end{eqnarray}
The \textbf{connected} $m$-point dimer correlation function instead involves derivatives of $\log {\cal Z}$ in the usual way:
\be
{\cal G}^{(m)}_c(\alpha_1,\ldots,\alpha_m) = \left(\prod_{k=1}^m \frac{\partial}{\partial \log w_{\alpha_k}}\right)\log {\cal Z}.
\ee
(The subscript $c$ on ${\cal G}_c^{(m)}$ means `connected' and should not be confused with the colour label introduced earlier.) 
These correlation functions are invariant under permutations of the $m$ edge indices $\alpha_1,\ldots,\alpha_m$. The correlation functions also depend on the
dimer weights of all edges in the system, but (just as for the generating function ${\cal Z}$) we do not indicate this dependence explicitly.

Next note that the rhs of (\ref{Zqcol}) is independent of the values of the dimer weights, so we can choose these at our convenience. By choosing to evaluate the rhs at $w_{\alpha}=1$, we can replace $\partial/\partial w_{\alpha}$ by $w_{\alpha}\partial/\partial w_{\alpha}=\partial/\partial \log w_{\alpha}$. Thus $Z$ can be rewritten as
\be
Z = \left(\prod_{\alpha}\frac{\partial}{\partial \log w_{\alpha}}\right){\cal Z}^q\Bigg|_{w_{\alpha}=1}.
\label{Zqcol2}
\ee
Starting from this expression, we show in Appendix \ref{proof} that $Z$ can be written as 
\be
Z = {\cal Z}^q \sum_{P=1}^{B_{|E|}}q^{N_{P}}\prod_{r=1}^{N_{P}}{\cal G}_c^{(m_{Pr})}(S_{Pr})\Bigg|_{w_{\alpha}=1}.
\label{Z-gf}
\ee
The sum is over the $B_{|E|}$ partitions of $E$ (here $B_{\ell}$ is the Bell number, the number of partitions of a set of  $\ell$ objects), and $N_P$ is the number of subsets in partition $P$. The integer $r=1,\ldots,N_P$ labels the subsets $S_{Pr}$ of the partition, $m_{Pr}$ is the number of edges in $S_{Pr}$, and ${\cal G}^{(m_{Pr})}_c(S_{Pr})$ is the connected dimer correlation function for this subset. 

\section{Fermionic (Grassmann integral) formulations}
\label{fermionic}

\begin{figure}[h]
\begin{centering}
\hspace{-6cm}
$
\begin{array}{c}
	\includegraphics[scale=0.23]{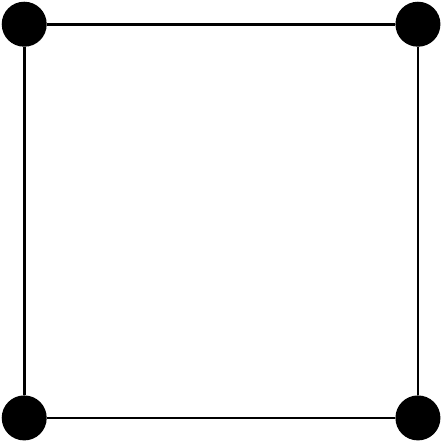} \\
	\mbox{(a)} \vspace{0.7cm}\\
	\includegraphics[scale=0.23]{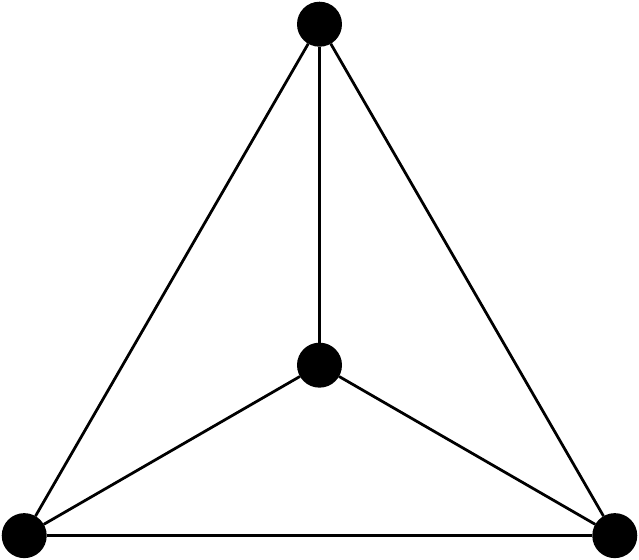} \\
	\mbox{(b)} 
\end{array}
$
\hspace{1cm}
\begin{minipage}{2in}
	$
	\begin{array}{ccc}
	\includegraphics[scale=0.24]{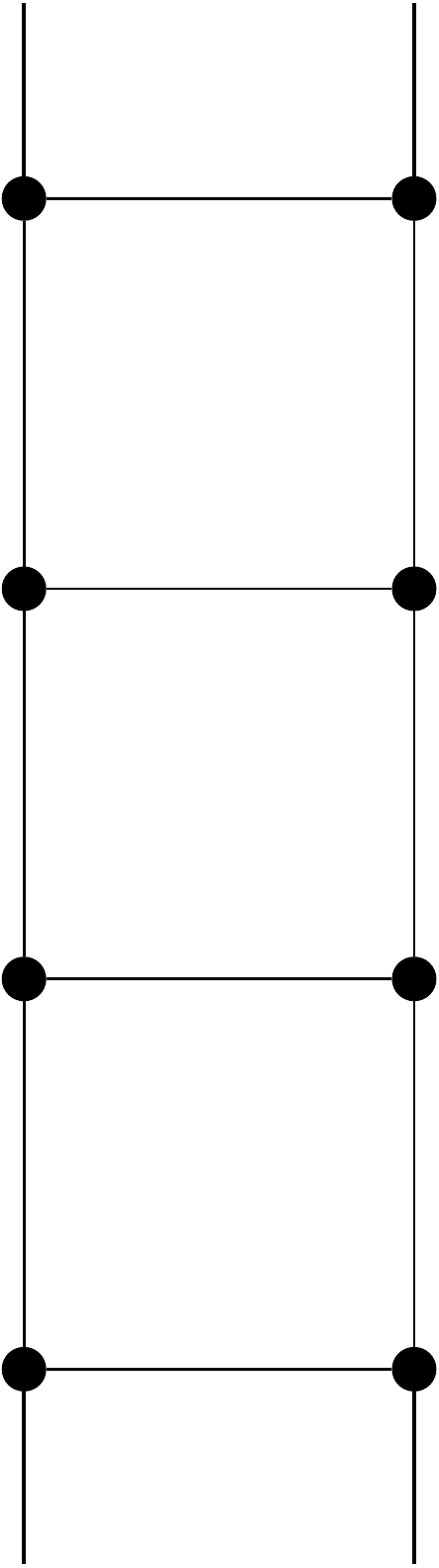} \hspace{1cm} & \includegraphics[scale=0.27]{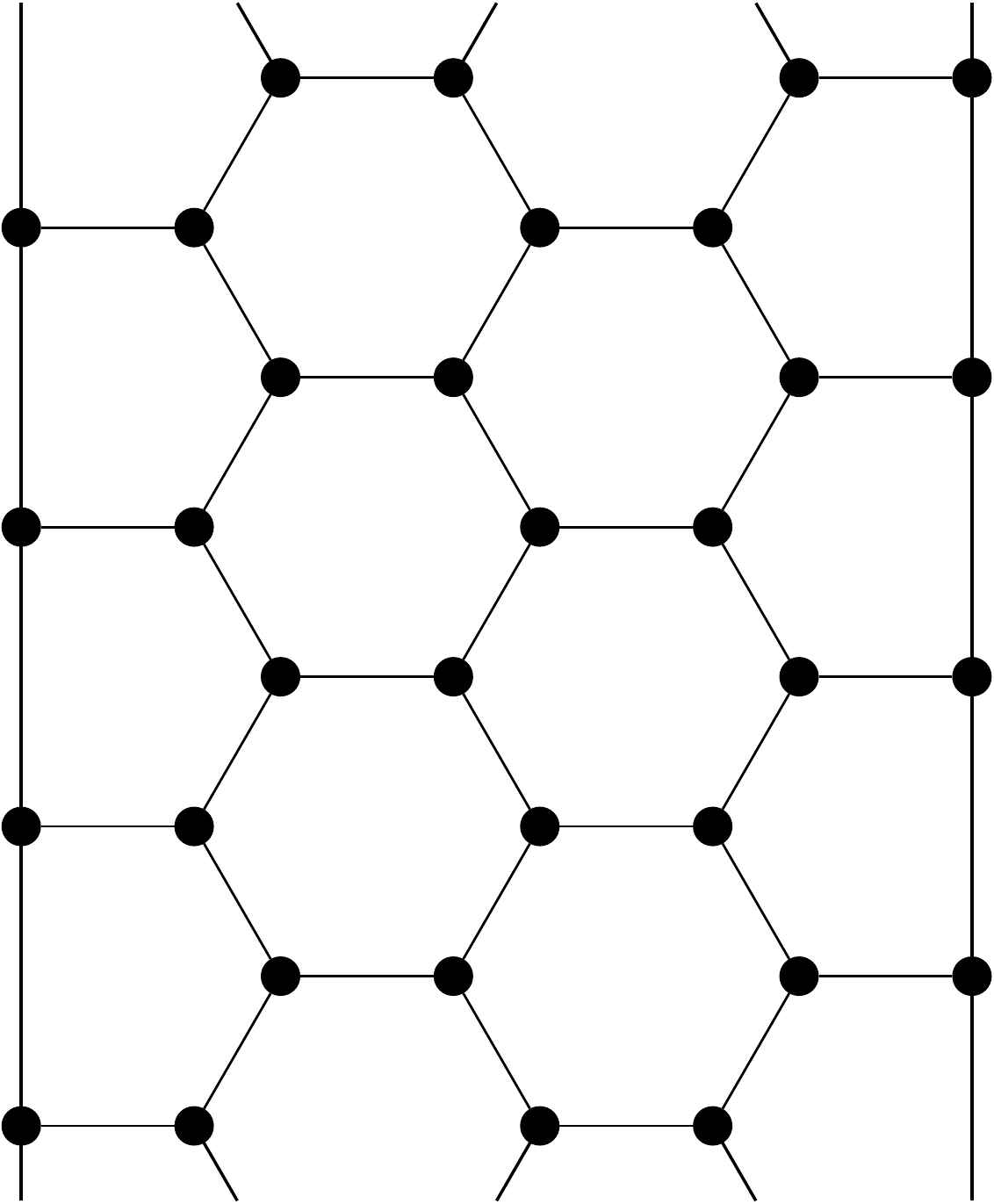}\hspace{1cm} & \includegraphics[scale=0.24]{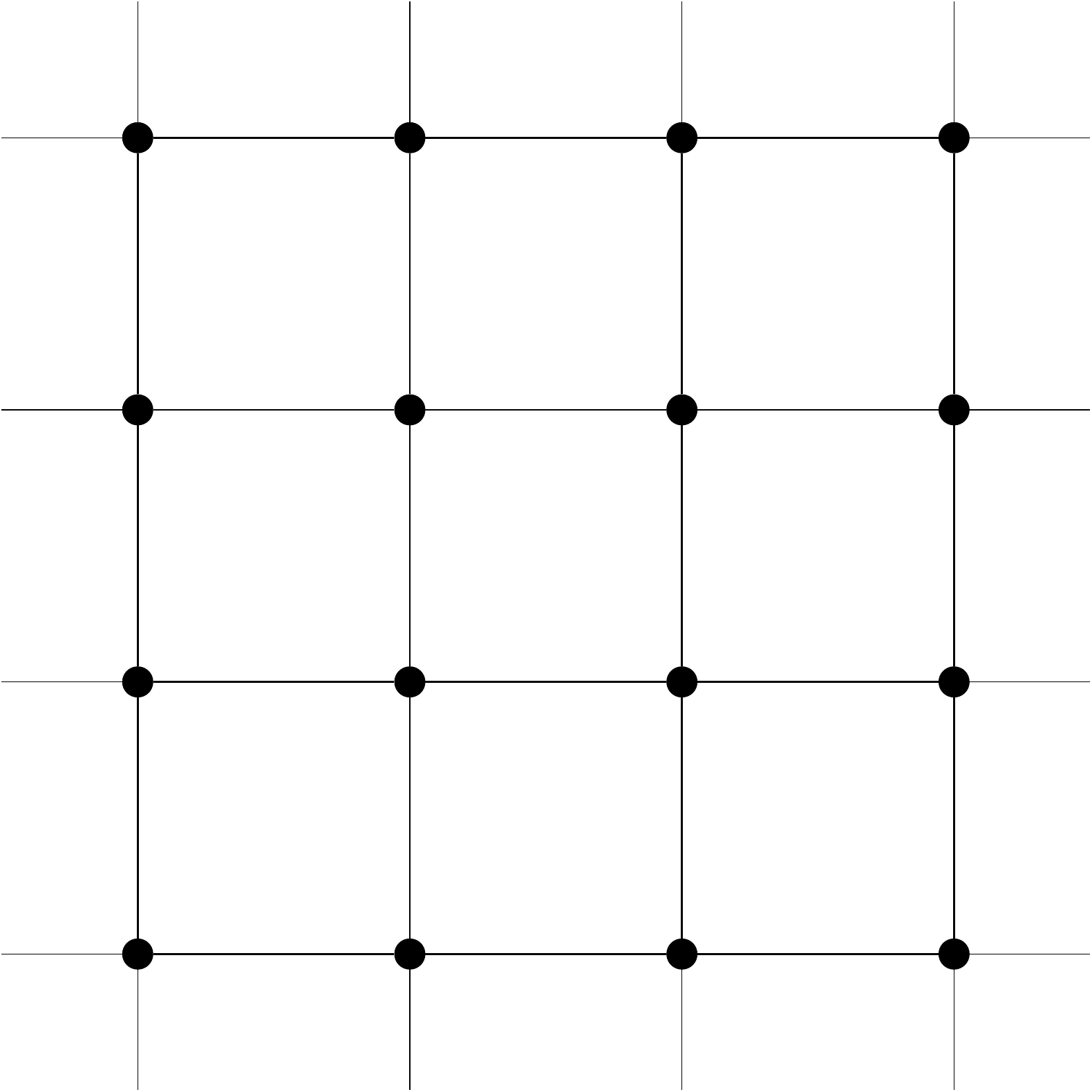} \\
	\hspace{-1cm}\mbox{(c)} & \hspace{-1cm}\mbox{(d)} & \mbox{(e)}  
	\end{array}
	$
\end{minipage}
\caption{Some examples of graphs of degree $q$ (with $q=2,3,4$) to which the methods developed in this paper can be applied to calculate the number of $q$-edge-colourings $Z$ (all graphs shown 
have $Z\neq 0$). In these figures, ``loose" edges on opposite sides of each of the two directions (horizontal and vertical) should be identified. Graphs (a)-(d) are planar, with (c)-(d) drawn here in terms of their alternative (and more natural) embedding on a cylinder, while (e) is embedded on a torus.}
\label{fig:graphs}
\end{centering}
\end{figure}

In this section we derive fermionic formulations, in the form of expressions involving Grassmann integrals, for the number of $q$-edge-colourings $Z$ of graphs of degree $q$ that can be embedded without edge crossings either on a plane (Sec. \ref{planar}) or on a torus (Sec. \ref{torus}). It is instructive first to consider some examples of such graphs, shown in Fig.~\ref{fig:graphs}. Graphs (a)-(d) are in the planar category while (e) is in the toroidal category. While the fermionic formulations are applicable to graphs of arbitrary size, including very small ones like graphs (a) and (b), regular lattice graphs of large (macroscopic) size are typically of greatest interest in statistical mechanics. Examples of such graphs in the toroidal category can be constructed very easily and naturally from Archimedean tilings by imposing periodic boundary conditions (BC's) in both directions. These tilings include many of the most frequently studied two-dimensional lattices in statistical mechanics, including the square lattice on which graph (e) is based. Graphs (c)-(d) are two examples of regular lattice graphs in the planar category that are most naturally embedded on a cylinder (rather than on a plane), which is also how they are drawn here (with open BC's in the horizontal direction and periodic BC's in the vertical direction). Graph (d) also exemplifies how a regular lattice graph in the planar category can be constructed by embedding an Archimedean tiling (in this case a honeycomb lattice) on a cylinder by imposing the appropriate BC's, and then introducing additional edges along the boundary (the vertical edges here) to ensure that the boundary vertices have the same degree as those in the bulk.

\subsection{The planar case}
\label{planar}

We start by considering a graph $G=(E,V)$ of degree $q$ as defined in Sec. \ref{dimerZ}. We may assume that the number of vertices $N$ is even, as otherwise the possibility of dimer coverings of $G$ (and in turn $q$-edge-colourings) is trivially ruled out. We introduce an orientation of $G$ defined by a set of arrows, one on each edge, each arrow pointing towards one of the two vertices connected to the edge. To a given orientation we associate a unique antisymmetric $N\times N$ matrix 
$M$ with matrix element $M_{ij}=0$ if vertices $i$ and $j$ are not connected by an edge, and $|M_{ij}|= 1$ if they are, with $M_{ij}=+1$ $(-1)$ if the arrow on the edge points from $i$ to $j$ (from $j$ to $i$). We will refer to $M$ as a signed adjacency matrix.

We next make a further restriction to planar graphs, which by definition can be embedded in the plane without crossing edges. From now on we consider such an embedding, which defines a set of faces. An orientation of the graph that has an odd number of edges oriented clockwise around each face will be called a Kasteleyn orientation (KO). Given a KO we can generate other KO's from it by $\mathbb{Z}_2$ gauge transformations, which conserve the clockwise-odd constraints. The elementary $\mathbb{Z}_2$ gauge transformation $\mathbb{G}_i$ reverses the arrow directions on all edges connected to vertex $i$. A general $\mathbb{Z}_2$ gauge transformation is then given by $\mathbb{G}_S=\prod_{i\in S}\mathbb{G}_i$ where $S\subset V$. Different KO's that are related by a $\mathbb{Z}_2$ gauge transformation are said to be gauge-equivalent. For planar graphs it turns out that all KO's are gauge-equivalent. 
Let $K$ denote
the signed adjacency matrix corresponding to a specific KO, and let $A$ be the antisymmetric $N\times N$ matrix defined by $A_{ij}=w_{ij}K_{ij}$, where $w_{ij}$ ($=w_{ji}$) is the edge-dependent dimer weight if $i$ and $j$ are connected by an edge, and zero otherwise. The dimer generating function is then given by \cite{kast-63}
\be
{\cal Z} = |\mbox{Pf }A|
\ee
where $\mbox{Pf}$ denotes the Pfaffian. As $\mathbb{G}_i$ changes $\mbox{Pf }A$ by a sign, $\mathbb{Z}_2$ gauge transformations leave ${\cal Z}$ invariant. If all dimer weights are set to unity, $A$ reduces to $K$ and ${\cal Z}$ becomes the number of dimer coverings. By introducing a Grassmann variable $\psi_i$ on each vertex $i$, $\mbox{Pf }A$ can be expressed as a Gaussian Grassmann integral over these variables:\footnote{An equivalent representation is $\mbox{Pf }A = \int d\psi_1\ldots d\psi_N\; \exp\left(-\frac{1}{2}\sum_{ij}\psi_i A_{ij}\psi_j\right)$.}
\be
\mbox{Pf }A = \int d\psi_N \ldots d\psi_1 \; \exp\left(\frac{1}{2}\sum_{ij}\psi_i A_{ij}\psi_j\right). 
\label{pf-grassmann}
\ee
Thus 
\be
{\cal Z}^q = \left| \int \left[\prod_{c=1}^q d\psi_N^{(c)}\ldots d\psi_1^{(c)}\right] \exp\left(\frac{1}{2}\sum_{c=1}^q \sum_{ij}\psi_i^{(c)}w_{ij}K_{ij}\psi_j^{(c)}\right)\right|
\label{Zqplanar}
\ee
where we introduced a colour superscript $c=1,\ldots,q$ to distinguish different Grassmann variables belonging to the same vertex. Using Eq. (\ref{Zqcol}) then gives  
\be
Z = \left|\int \left[\prod_{c=1}^q d\psi_N^{(c)}\ldots d\psi_1^{(c)}\right] \exp\left(\frac{1}{2}\sum_{c=1}^q \sum_{ij}\psi_i^{(c)}w_{ij}K_{ij}\psi_j^{(c)}\right)  
\prod_{\alpha}\sum_{c_{\alpha}=1}^q \psi_{i_{1\alpha}}^{(c_{\alpha})}K_{i_{1\alpha},i_{2\alpha}}\psi_{i_{2\alpha}}^{(c_{\alpha})}\right|.
\label{Z-prelim1}
\ee
(In the product over edges, we have labeled the two vertices touching an edge $\alpha$ as $i_{1\alpha}$ and $i_{2\alpha}$. Thus in this product, a given
Grassmann variable $\psi_i^{(c)}$ will in fact appear under $q$ different names, one for each edge the vertex $i$ touches.) This expression can be simplified further. First, note that expanding out the product over edges gives a sum of terms, each of which contains $2|E|=qN$ Grassmann variables (not necessarily all distinct), which also equals the total number of Grassmann variables being integrated over. From the properties of Grassmann variables it then follows that the exponential can be replaced by 1. This shows that the rhs of Eq. (\ref{Z-prelim1}) is independent of the dimer weights $w_{\alpha}$. Second, the signs $K_{i_{1\alpha},i_{2\alpha}}$ in the product over edges are superfluous and can be omitted. 
Thus
\be
Z = \left|\int \left[\prod_{c=1}^q d\psi_N^{(c)}\ldots d\psi_1^{(c)}\right] \prod_{\alpha}\sum_{c_{\alpha}=1}^q \psi_{i_{1\alpha}}^{(c_{\alpha})}\psi_{i_{2\alpha}}^{(c_{\alpha})}\right|.
\label{Z-main-1}
\ee
Next we introduce a pair of Grassmann variables $\bar{\xi}_{\alpha}$ and $\xi_{\alpha}$ on each edge $\alpha$ (for an edge connecting vertices $i$ and $j$ we can write the variables as $\bar{\xi}_{ij}=\bar{\xi}_{ji}$ and $\xi_{ij}=\xi_{ji}$). 
Then $Z$ can be written as
\be
Z = \left|\int \left[\prod_{c=1}^q d\psi_N^{(c)}\ldots d\psi_1^{(c)}\right]\left[\prod_{\alpha}d\bar{\xi}_{\alpha}d\xi_{\alpha}\right] 
\exp\left(\frac{1}{2}\sum_{c=1}^q \sum_{ij}\psi_i^{(c)}M_{ij}\bar{\xi}_{ij}\xi_{ij}\psi_j^{(c)}\right)\right|.
\label{Z-main-2}
\ee
Here $M$ is the signed adjacency matrix corresponding to an \textit{arbitrary} orientation of the graph. 
To prove this result one simply integrates out the edge Grassmann variables, which leads back to Eq. (\ref{Z-main-1}). 

Eqs. (\ref{Z-main-1}) and (\ref{Z-main-2}) are the main results of this subsection. Although Eq. (\ref{Z-main-2}) does in some sense represent a less economical formulation than Eq. (\ref{Z-main-1}) due to the additional
Grassmann variables living on the edges, Eq. (\ref{Z-main-2}) is of interest because it takes the ``standard" form $\int[$integration measure$]\exp($action$)$ 
for a partition function, with the action a sum of spatially local terms; this form may be more convenient for further manipulations. The fact that the action in Eq. (\ref{Z-main-2}) is quartic in the Grassmann variables, i.e. non-Gaussian, 
is a reflection of the interacting nature of the problem. We also emphasize that since $M$ in Eq. (\ref{Z-main-2}) can be any signed adjacency matrix for the graph, Eq. (\ref{Z-main-2}) for $Z$ is invariant under reversals of the 
orientations of an arbitrary subset of edges. This is a much larger set of transformations than
the $\mathbb{Z}_2$ gauge transformations that leave the dimer generating function ${\cal Z}$ invariant.

The validity of Eq. (\ref{Z-main-2}) can be verified using a somewhat different line of reasoning, starting from the observation (cf. the discussion in Sec.~\ref{dimerZ}) that the calculation of the number of $q$-edge-colourings $Z$ 
is similar to the calculation of the number of $q$-dimer-coverings ${\cal Z}^q|_{w_{\alpha}=1}$, but with the crucial difference that all $q$-dimer-coverings that contain edges with multiple dimers are not included in the count.  
This suggests that in order to find $Z$ one can first modify Eq. (\ref{Zqplanar}) (with all dimer weights set to unity) by making the replacement $K_{ij}\to \widetilde{K}_{ij}=K_{ij}\bar{\xi}_{ij}\xi_{ij}$. We can think of the 
$\widetilde{K}_{ij}$ as matrix elements of a ``Grassmann-valued Kasteleyn matrix". As these matrix elements contain a product of an even number of Grassmann variables, they behave like ordinary $c$-numbers (in particular, they commute with everything) except that their square vanishes by virtue of $\bar{\xi}_{ij}^2=\xi_{ij}^2 = 0$. This latter property ensures that all $q$-dimer-coverings containing edges with multiple dimers will give zero contribution. Integrating over the vertex Grassmann variables therefore gives $|Z \prod_{\alpha}\bar{\xi}_{\alpha}\xi_{\alpha}|$. Introducing also an integral over the edge Grassmann variables, one is left simply with $Z$. Finally, one can convince oneself that the same final result is obtained if $K$ is replaced by an arbitrary signed adjacency matrix $M$ for the graph.

\subsection{The toroidal case}
\label{torus}

Here we consider nonplanar graphs of degree $q$ with the same defining properties as those discussed in Sec. \ref{planar}, except that these graphs can be embedded on a torus, not on a plane (without crossing edges). 
As before, an orientation is described by a signed adjacency matrix $M$. Kasteleyn orientations and $\mathbb{Z}_2$ gauge transformations are also defined in the same way as for planar graphs. For toroidal graphs the Kasteleyn orientations fall into 4 gauge-inequivalent classes. 
The dimer generating function ${\cal Z}$ can be written
\be
{\cal Z} = \frac{1}{2}\left|\sum_{\mu=1}^4 r_{\mu}\mbox{Pf }A^{(\mu)}\right|,
\label{dimZtorus}
\ee
where the sum goes over the 4 gauge-inequivalent classes, $A^{(\mu)}_{ij} = w_{ij}K_{ij}^{(\mu)}$, where $K^{(\mu)}$ is the Kasteleyn matrix corresponding to 
a Kasteleyn orientation chosen from class $\mu$, and the $r_{\mu}=\pm 1$ are appropriately chosen signs. We refer to the literature for more detailed discussions of these matrices and signs 
\cite{kast-61,dolbilin,regge,galluccio,tesler,cimasoni}. 

Following the same steps as in Sec. \ref{planar}, we now find  
\be
{\cal Z}^q = 2^{-q}\left|\sum_{\mu_1,\ldots,\mu_q} r_{\mu_1}\cdots r_{\mu_q}\int \left[\prod_{c=1}^q d\psi_N^{(c)}\ldots d\psi_1^{(c)}\right]
\exp\left(\frac{1}{2}\sum_{c=1}^q\sum_{ij}\psi_i^{(c)}w_{ij}K_{ij}^{(\mu_c)}\psi_j^{(c)}\right)\right|
\ee
and
\begin{eqnarray}
Z &=& 2^{-q}\Bigg|\sum_{\mu_1,\ldots,\mu_q} r_{\mu_1}\cdots r_{\mu_q}\int \left[\prod_{c=1}^q d\psi_N^{(c)}\ldots d\psi_1^{(c)}\right] 
\exp\left(\frac{1}{2}\sum_{c=1}^q\sum_{ij}\psi_i^{(c)}w_{ij}K_{ij}^{(\mu_c)}\psi_j^{(c)}\right)\nonumber \\ & \times &
\prod_{\alpha}\sum_{c_{\alpha}=1}^q \psi_{i_{1\alpha}}^{(c_{\alpha})}K_{i_{1\alpha},i_{2\alpha}}^{(\mu_{c_{\alpha}})}\psi_{i_{2\alpha}}\Bigg|.
\end{eqnarray}
The exponential can be replaced by 1 by the same argument as for the planar case. The signs contributed by the signed adjacency matrices can however not be completely
eliminated, unlike the situation for the planar case. But $Z$ is still invariant under reversal of the orientations of an arbitrary subset of edges. As noted before, this set of transformations is much larger than the $\mathbb{Z}_2$ gauge transformations, and thus the signed adjacency matrices $M^{(\mu)}$ resulting
from these transformations will generally not satisfy clockwise-odd constraints around the faces. Note however that the change in $M^{(\mu)}$ is independent of $\mu$ and 
thus the transformations preserve the ratios $M^{(\mu)}_{ij}/M^{(\mu')}_{ij}$. We get
\be
Z = 2^{-q}\Bigg|\sum_{\mu_1,\ldots,\mu_q} r_{\mu_1}\cdots r_{\mu_q}\int \left[\prod_{c=1}^q d\psi_N^{(c)}\ldots d\psi_1^{(c)}\right] 
\prod_{\alpha}\sum_{c_{\alpha}=1}^q \psi_{i_{1\alpha}}^{(c_{\alpha})}M_{i_{1\alpha},i_{2\alpha}}^{(\mu_{c_{\alpha}})}\psi_{i_{2\alpha}}\Bigg|.
\label{Z-t-main-1}
\ee
Note that $M^{(\mu)}_{i_{1\alpha},i_{2\alpha}}$ is independent of $\mu$ for some subset of the edges. For these edges this sign is superfluous in this expression and may be omitted.

Next, by introducing Grassmann variables $\bar{\xi}_{\alpha}$ and $\xi_{\alpha}$ on each edge $\alpha$, 
we can express $Z$ as 
\begin{eqnarray} 
Z &=& 2^{-q}\Bigg|\sum_{\mu_1,\ldots,\mu_q} r_{\mu_1}\cdots r_{\mu_q} \nonumber \\ & \times & 
\int \left[\prod_{c=1}^q d\psi_N^{(c)}\ldots d\psi_1^{(c)}\right] \left[\prod_{\alpha}d\bar{\xi}_{\alpha}d\xi_{\alpha}\right]
\exp\left(\frac{1}{2}\sum_{c=1}^q\sum_{ij}\psi_{i}^{(c)}M_{ij}^{(\mu_{c})} \bar{\xi}_{ij}\xi_{ij}\psi_{j}^{(c)}\right)\Bigg|.
\label{Z-t-main-2}
\end{eqnarray}
Eqs. (\ref{Z-t-main-1}) and (\ref{Z-t-main-2}) are the main results of this subsection. They are the torus analogues of Eqs. (\ref{Z-main-1}) and (\ref{Z-main-2}) for the planar case. 
When properly generalized, the remarks made at the end of Sec. \ref{planar} are valid also for Eqs. (\ref{Z-t-main-1}) and (\ref{Z-t-main-2}). The only difference between the two cases is that in the toroidal
case the analysis involves four different signed adjacency matrices $M^{(\mu)}$ instead of one, which leads to a sum of $4^q$ terms in the expression for $Z$. 
The number of distinct terms is however less than $4^q$ due to the invariance of the summands in Eqs. (\ref{Z-t-main-1}) 
and (\ref{Z-t-main-2}) under permutations of $\mu_1,\ldots,\mu_q$. Moreover, as we will see an example of in the next section, additional simplifications may be possible which
can allow one to further reduce, to a handful or less, the number of terms that need to be explicitly evaluated.

\section{Exact numerical evaluation of Grassmann integral expressions: application to 4-edge colourings on a square lattice embedded on a torus}
\label{num}

\begin{table}[!htb]
 \begin{center}
\noindent 
\scriptsize
\begin{tabular}{|c|r||c|r|}\hline
 $N_x \times N_y$ & $Z$ & $N_x \times N_y$ & $Z$ \\\hline 
 $3\times 4$ & 1\,440 & $5\times 24$ &  75\,493\,427\,239\,878\,525\,921\,802\,080 \\
 $3\times 6$ & 14\,688 & $5\times 26$ &  9\,101\,986\,590\,192\,512\,415\,402\,888\,960 \\
 $3\times 8$ & 168\,480 & $5\times 28$ &  1\,097\,406\,776\,113\,872\,427\,375\,721\,635\,200 \\
 $3\times 10$ & 1\,998\,432 & $5\times 30$ &  132\,312\,447\,510\,919\,624\,125\,369\,146\,179\,680 \\ \cline{3-4}
 $3\times 12$ &  23\,911\,200 & $6\times 6$ &  1\,072\,652\,544 \\        
 $3\times 14$ &  286\,724\,448 & $6\times 7$ &  15\,664\,050\,528 \\ 
 $3\times 16$ &  3\,440\,063\,520 & $6\times 8$ &  589\,685\,031\,168 \\
 $3\times 18$ &  41\,278\,872\,672 & $6\times 9$ &  12\,978\,406\,493\,280 \\  
 $3\times 20$ &  495\,340\,803\,360 & $6\times 10$ &  412\,755\,112\,206\,336 \\ 
 $3\times 22$ &  5\,944\,072\,634\,208 & $6\times 11$ &  10\,668\,372\,957\,077\,088 \\  
 $3\times 24$ &  71\,328\,820\,592\,160 & $6\times 12$ & 318\,152\,750\,518\,821\,888 \\  
 $3\times 26$ &  855\,945\,694\,050\,912 & $6\times 13$ &  8\,758\,404\,262\,969\,876\,320 \\        
 $3\times 28$ & 10\,271\,347\,869\,445\,920 & $6\times 14$ &  254\,822\,001\,471\,158\,993\,664 \\  \cline{1-2}
 $4 \times 4$ & 44\,160 & $6\times 15$ &  7\,188\,925\,477\,697\,173\,731\,168 \\       
 $4 \times 5$ & 243\,840 & $6\times 16$ &  207\,182\,876\,960\,104\,821\,689\,088 \\ 
 $4 \times 6$ & 3\,629\,568 & $6\times 17$ &  5\,900\,501\,794\,802\,890\,396\,061\,280 \\  
 $4 \times 7$ & 35\,750\,400 &  $6\times 18$ & 169\,431\,182\,476\,119\,228\,309\,023\,232 \\ 
 $4 \times 8$ & 451\,209\,216 & $6\times 19$ &  4\,842\,968\,738\,352\,136\,409\,208\,425\,568 \\  
 $4 \times 9$ & 5\,158\,471\,680 & $6\times 20$ & 138\,869\,108\,263\,890\,882\,603\,021\,792\,768 \\ \cline{3-4} 
 $4 \times 10$ & 62\,658\,256\,896 & $7\times 8$ & 11\,607\,147\,272\,448 \\  
 $4\times 11$ & 742\,987\,407\,360 & $7\times 10$ & 10\,486\,277\,124\,715\,392 \\ 
 $4\times 12$ & 8\,942\,452\,899\,840 & $7\times 12$ & 10\,237\,845\,291\,845\,195\,808 \\ 
 $4\times 13$ & 106\,992\,869\,867\,520 & $7\times 14$ & 10\,322\,131\,976\,601\,474\,883\,392 \\ 
 $4\times 14$ & 1\,284\,862\,426\,349\,568 & $7\times 16$ & 10\,552\,848\,259\,199\,993\,065\,837\,056 \\ 
 $4\times 15$ & 15\,407\,016\,206\,008\,320 & $7\times 18$ & 10\,855\,518\,364\,633\,186\,870\,098\,974\,592 \\ 
 $4\times 16$ & 184\,918\,168\,859\,836\,416 & $7\times 20$ & 11\,198\,031\,662\,235\,847\,304\,452\,806\,203\,520 \\ 
 $4\times 17$ &  2\,218\,611\,020\,841\,615\,360 &  $7\times 22$ & 11\,566\,056\,629\,691\,121\,682\,663\,125\,245\,340\,416 \\ 
 $4\times 18$ &  26\,624\,552\,900\,185\,030\,656 & $7\times 24$ & 11\,953\,167\,548\,050\,678\,272\,222\,497\,989\,683\,044\,256 \\ 
 $4\times 19$ &  319\,479\,997\,996\,235\,489\,280 & $7\times 26$ & 12\,356\,574\,891\,854\,133\,626\,297\,339\,854\,605\,902\,125\,248 \\ 
 $4\times 20$ &  3\,833\,803\,880\,694\,043\,115\,520 & $7\times 28$ & 12\,775\,197\,930\,055\,212\,382\,651\,748\,538\,403\,428\,081\,779\,328 \\  \cline{1-2}
 $5 \times 6$ & 17\,988\,960 & $7\times 30$ & 13\,208\,772\,604\,211\,970\,774\,339\,193\,141\,679\,044\,412\,032\,635\,168 \\ 
 $5 \times 8$ & 1\,840\,646\,400 & $7\times 32$ & 13\,657\,432\,485\,307\,219\,457\,227\,579\,236\,006\,517\,304\,339\,033\,098\,752 \\ \cline{3-4}
 $5 \times 10$ &  209\,977\,817\,280 & $8\times 8$ & 1\,685\,928\,423\,086\,592 \\
 $5\times 12$ &  24\,836\,803\,964\,640 & $8\times 9$ & 65\,640\,522\,173\,916\,672 \\ 
 $5\times 14$ &  2\,974\,334\,794\,053\,120 & $8\times 11$ & 361\,943\,772\,232\,276\,810\,752 \\
 $5\times 16$ &  357\,739\,836\,702\,854\,400 & $8\times 13$ & 1\,983\,974\,100\,428\,341\,796\,935\,680 \\
 $5\times 18$ &  43\,094\,084\,170\,133\,825\,760 & $8\times 15$ & 10\,860\,078\,679\,370\,714\,987\,464\,273\,920 \\ 
 $5\times 20$ &  5\,194\,113\,814\,423\,956\,157\,440 & $8\times 17$ & 59\,428\,139\,978\,826\,027\,050\,009\,652\,486\,144 \\
 $5\times 22$ &  626\,172\,572\,098\,389\,717\,579\,840 & $8\times 19$ & 325\,176\,895\,253\,864\,837\,579\,133\,026\,112\,749\,568  \\\hline 
  \end{tabular}
 \normalsize
 \end{center}
  \caption{\label{tab:numeval}The number of 4-edge-colourings $Z=Z(N_x,N_y)$ on a square lattice of size $N_x \times N_y$ embedded on a torus. The results were obtained by exact numerical evaluation of Eq. (\ref{Z-t-main-2}), 
 using the algorithm in \protect\cite{creutz} to evaluate the Grassmann integrals. Only cases with  $N_x\leq N_y$ are shown, as $Z$ is invariant under interchange of $N_x$ and $N_y$. (For $N_x=8$ with $N_y>9$ only 
 odd values of $N_y$ were considered, as the computation times for even values of $N_y$ were prohibitively large.)}
 \end{table}

\begin{table}[!htb]
 \begin{center}
\noindent 
\begin{tabular}{|c||c|c|c|c|}\hline
 $N_x$ & $d_{\rm{max}}$ & $d_{\rm{min}}$ & $f(N_x,\infty)$ & $\lambda_{\rm{max}}$ \\\hline
 3 & 4 & 4 & 0.41415 & 3.4641 \\
 4 & 1 & 0 & 0.62123 & 12.000 \\
 5 & 4 & 4 & 0.47922 & 10.980 \\
 6 & 1 & 0 & 0.55919 & 28.649 \\
 7 & 4 & 4 & 0.49580 & 32.156 \\
 8 & 1 & 0 & 0.53796 & 73.971 \\\hline
\end{tabular}
\end{center}
\caption{\label{tab:extracted} Values of $d_{\rm{max}}$, $d_{\rm{min}}$, $f(N_x,\infty)$, and $\lambda_{\rm{max}}=\exp(N_x f(N_x,\infty))$ obtained by fitting $f(N_x,N_y)$ in Fig. \ref{fig:4col-plots}(a) to the expression for $f(N_x,N_y)$ predicted by the transfer matrix approach for large $N_y$. For $N_x = 3$--$6$ we have verified the values listed here by also finding $\lambda_{\rm{max}}$, $d_{\rm{max}}$, and $d_{\rm{min}}$ directly from numerical diagonalization of the transfer matrix.}
\end{table}

In this section we present an approach for explicit calculation of the number of $q$-edge-colourings $Z$ on graphs of degree $q$, based on numerically
evaluating expressions for $Z$ like Eqs. (\ref{Z-main-2}) or (\ref{Z-t-main-2}). These involve Grassmann integrals on the form $\int d\eta_n \ldots d\eta_1\exp(S(\{\eta\}))$ which can 
be evaluated using an algorithm due to Creutz \cite{creutz}. 

As a concrete example of this approach, we consider the 4-edge-colouring problem on a toroidal square lattice (cf. Fig. \ref{fig:graphs}(e)) with $N_x$ ($N_y$) vertices in the $x$ ($y$) direction. Thus 
$Z$ can be found from Eq. (\ref{Z-t-main-2}), with appropriate choices for the matrices $M^{(\mu)}$ and the signs $r_{\mu}$ obtained from the Pfaffian solution of the dimer problem on this lattice \cite{kast-61}. 
While Eq. (\ref{Z-t-main-2}) in this case involves a sum of $4^4=256$ Grassmann integrals, the invariance of the summand under permutations of $\mu_1,\ldots,\mu_4$ can be used to reduce the number of terms to 35. 

Some aspects of our implementation of Creutz's algorithm are discussed in Appendix \ref{impl}. The nature of the algorithm and the specific problem to which it is applied, together with generic limitations on computer speed and memory, imply that in practice the maximum value of $N_x$  that can be considered is rather small; in our implementation we were able to go up to $N_x=8$. The maximum values of $N_y$ can be considerably larger but decrease as a function of $N_x$. 
For all calculations done with $N_x=3$--$6$ we evaluated all 35 terms and observed that many of them evaluated to zero while many others cancelled among themselves. As a 
consequence the correct answer for $Z$ could in fact be obtained by restricting evaluation to a much smaller number of terms: 5 terms if both $N_x$ and $N_y$ are even, and only 1 term
if either $N_x$ or $N_y$ is odd. This made it possible for us to also calculate $Z$ for cases with $N_x=7$ and 8 that otherwise would have been too time-consuming. Table
\ref{tab:numeval} shows our results for $Z=Z(N_x,N_y)$ for $N_x =$ 3--8. Note that for $N_x=8$, we restricted the largest $N_y$ values to be odd, as this allowed us to find $Z$ by calculating only one Grassmann integral.

To check the correctness of the results in Table \ref{tab:numeval} we have compared them in various ways to results from numerical transfer matrix calculations and also to a previous Bethe ansatz study \cite{dc-nien-04}. To this end, we write $Z$ as 
\be
Z=\mbox{Tr } T^{N_y}=\sum_i \lambda_i^{N_y}
\label{Ztm}
\ee
where the sum is over the eigenvectors $i$ of the $4^{N_x}\times 4^{N_x}$ row transfer matrix $T$ with eigenvalues $\lambda_i$. We have diagonalized $T$ numerically for $N_x=3$--$6$ and verified 
that evaluating $Z$ from Eq. (\ref{Ztm}) reproduces the results in Table \ref{tab:numeval} for not too large values of $N_y$ (this maximum value of $N_y$ decreases with $N_x$; for larger values of $N_y$ computational
errors due to the floating-point arithmetic involved in the transfer matrix calculation of $Z$ become too large). Our results for $N_x=7, 8$ with $N_y\leq 6$ are also indirectly verified in this way due to the symmetry $Z(N_x,N_y)=Z(N_y,N_x)$. 

\begin{figure}[h]
\hspace{-0.25cm}
\includegraphics[scale=0.67]{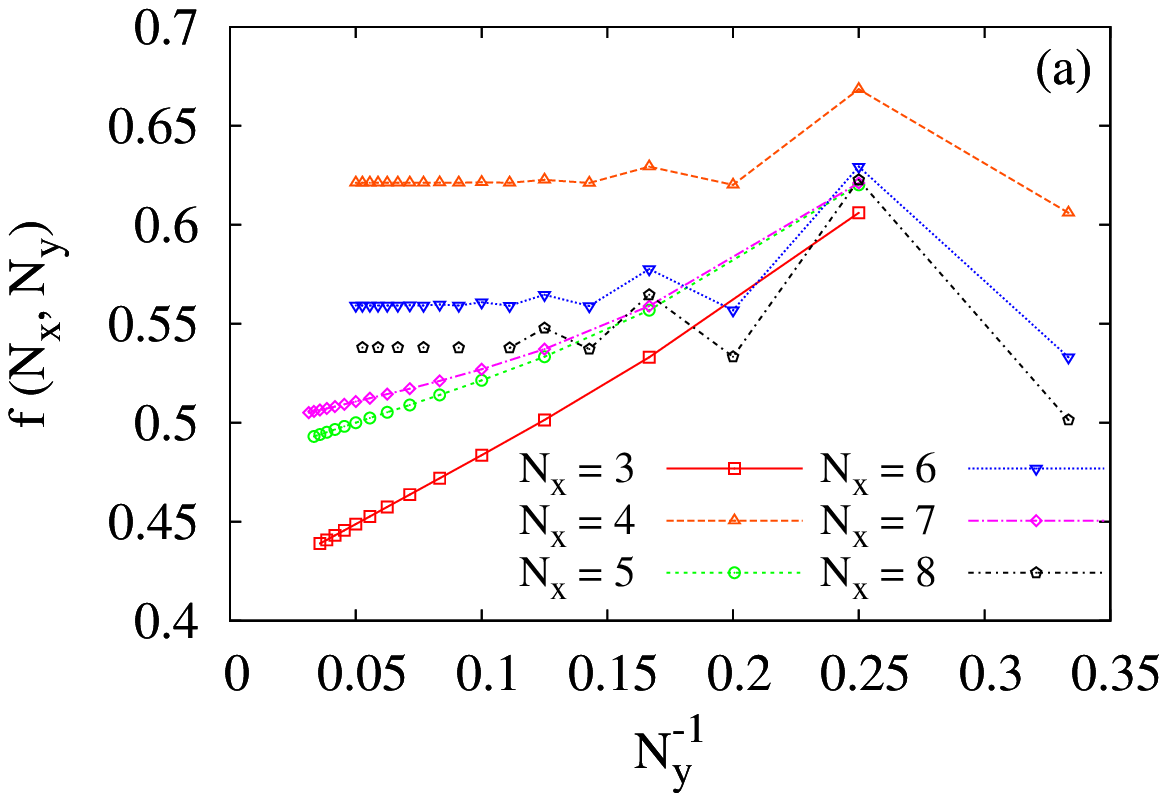}
\includegraphics[scale=0.67]{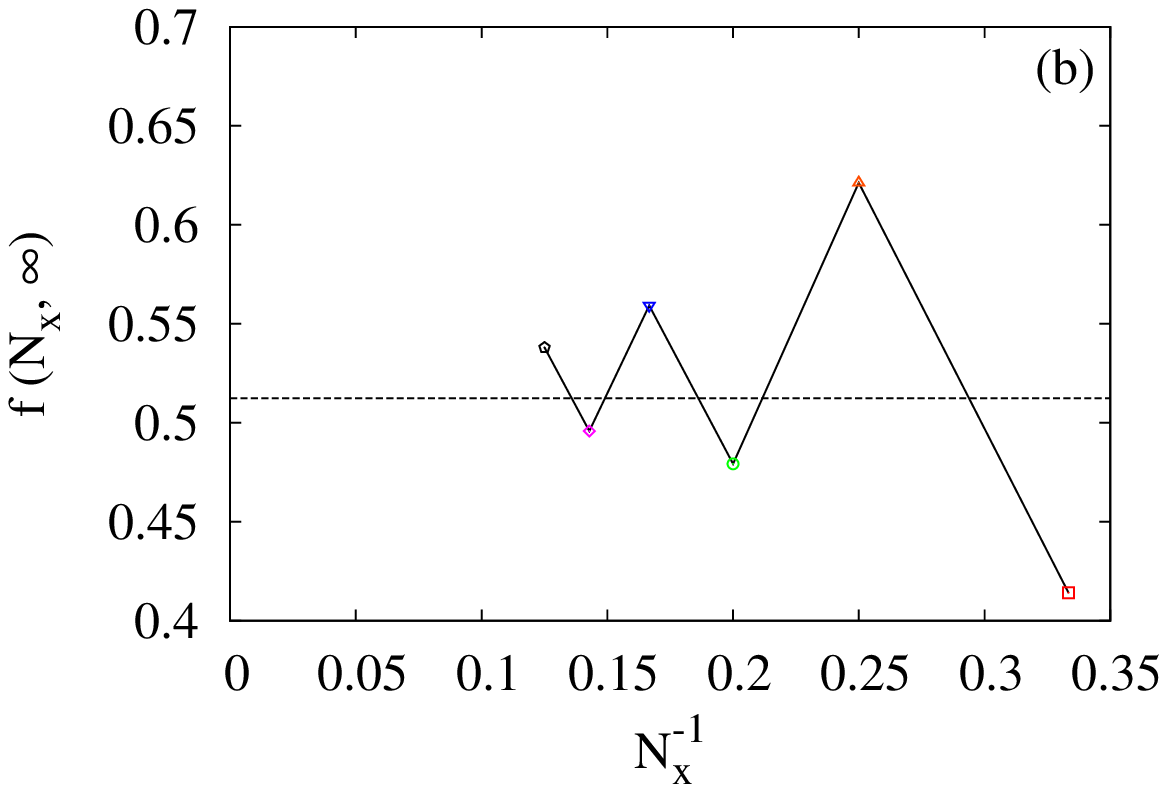}
\caption{(a) $f(N_x,N_y)\equiv (N_x N_y)^{-1}\log Z(N_x,N_y)$ plotted as a function of $N_y^{-1}$ for fixed values of $N_x$, with $Z$ obtained from Table \ref{tab:numeval}. 
(b) $f(N_x,\infty)$ as a function of $N_x^{-1}$ (symbols are the same as in (a)). The values of $f(N_x,\infty)$ (listed in Table \ref{tab:extracted}) were obtained as the intercepts of the extrapolation of $f(N_x,N_y)$ in (a) with the vertical axis $N_y^{-1}=0$. 
As $N_x^{-1}\to 0$ we expect $f(N_x,\infty)$ to approach the Bethe ansatz prediction \protect\cite{dc-nien-04} for $f(\infty,\infty$) (the horizontal dashed line). In both (a) and (b), the straight lines connecting points are guides to the eye. 
(Such lines were not drawn for the $N_x=8$ case when $N_y > 9$ as in this parameter regime $Z$ was not calculated for even values of $N_y$; cf. remark in caption of Table \ref{tab:numeval}.)}
\label{fig:4col-plots}
\end{figure}

For our further checks it is useful to consider $f(N_x,N_y)\equiv(N_x N_y)^{-1}\log Z(N_x,N_y)$ which in Fig. \ref{fig:4col-plots}(a) has been plotted as a function of $N_y^{-1}$ for fixed values
of $N_x$ (with $N_y$ restricted to values such that $N_x N_y$ is even, as $Z=0$ otherwise). The oscillating behaviour of $f$ for even $N_x$ can be understood as a consequence of negative
eigenvalues of $T$, as these contribute to $Z$ with opposite signs for even and odd $N_y$. As $N_y^{-1}$ becomes small, $f$ is seen to approach a straight line in all cases. 
To understand this we first note that for large $N_y$ the sum in Eq. (\ref{Ztm}) is dominated by the terms involving the eigenvalues with largest modulus, i.e. $Z\sim \lambda_{\rm{max}}^{N_y}[d_{\rm{max}} + d_{\rm{min}} (-1)^{N_y}]$, where $\lambda_{\rm{max}}>0$ is the largest eigenvalue, $d_{\rm{max}}$ is its multiplicity, and $d_{\rm{min}}$ is the multiplicity of the eigenvalue $-\lambda_{\rm{max}}$. Thus in this limit 
$f(N_x,N_y) \approx N_x^{-1}\log \lambda_{\rm{max}} +N_y^{-1} N_x^{-1}\log[d_{\rm{max}}+(-1)^{N_y}d_{\rm{min}}]$. By fitting this expression to the small-$N_y^{-1}$ region of the 
curves in Fig. \ref{fig:4col-plots}(a) one can deduce the values of $f(N_x,\infty)$, $\lambda_{\rm{max}}$, $d_{\rm{max}}$, and $d_{\rm{min}}$. The results are given in Table \ref{tab:extracted}. 
Our numerical diagonalization of $T$ for $N_x=3$--$6$ reproduces these values for those cases. 
 
We briefly discuss how the fitting is done. First consider the case of odd $N_x$, for which the slope of the straight line is $N_x^{-1}\log[d_{\rm{max}}+d_{\rm{min}}]$. Since $d_{\rm{max}}+d_{\rm{min}}$ must be an integer, its value
can be deduced with certainty from the slope, and in turn the exact value of the slope can be found. Next one can estimate the intercept $f(N_x,\infty)$ with the vertical axis, from which one gets   
$\lambda_{\rm{max}}=\exp(N_x f(N_x,\infty))$. Finally, since $Z=0$ for odd $N_y$ it follows from the large-$N_y$ expression for $Z$ in this case that $d_{\rm{min}}=d_{\rm{max}}$. For the case
of even $N_x$ the deduction of the multiplicities is slightly different. From Fig. \ref{fig:4col-plots} one sees that in this case $f$ becomes increasingly independent of $N_y^{-1}$ as $N_y^{-1}$ is decreased. Comparing this behaviour to the
coefficient $N_x^{-1}\log[d_{\rm{max}}+(-1)^{N_y}d_{\rm{min}}]$ of the $N_y^{-1}$ term in $f$, one is able to conclude that $d_{\rm{max}}=1$ and $d_{\rm{min}}=0$ (and so the slope is identically zero).

Finally, we show in Fig. \ref{fig:4col-plots}(b) $f(N_x,\infty)$ as a function of $N_x^{-1}$ together with the Bethe ansatz prediction \cite{dc-nien-04} 
for $f(\infty,\infty)$ (horizontal dashed line), given by $\log(\Gamma^2(1/4)/[\sqrt{2}\,\pi^{3/2}]) \approx 0.51238$. The observed behaviour of $f(N_x,\infty)$ as $N_x^{-1}$ decreases  
is clearly consistent with this prediction.

\section{Concluding remarks}
\label{remarks}

We have shown that the number $Z$ of $q$-edge-colourings of simple regular graphs of degree $q$ are deducible from the dimer generating function ${\cal Z}$ (or, equivalently, from the
set of connected dimer correlation functions) on the same graph. For graphs of this type that are either planar or embeddable on a torus we invoked the expressions for ${\cal Z}$ in terms of Pfaffians of Kasteleyn matrices 
to derive fermionic expressions for $Z$ in the form of Grassmann integrals. In particular, we derived expressions (Eqs. (\ref{Z-main-2}) and (\ref{Z-t-main-2})) in which the integrands are given as 
the exponential of a non-Gaussian quartic ``action" that is a sum of spatially local terms involving both vertex and edge Grassmann variables. 

As an example of a possible use of these expressions, we discussed exact numerical evaluations of them using an algorithm by Creutz \cite{creutz}, and presented  
a concrete application to the enumeration of $4$-edge-colourings on a square lattice embedded on a torus. The agreement found with results from Bethe ansatz \cite{dc-nien-04} and numerical transfer matrix calculations 
serves as a direct nontrivial check of the correctness of Eq. (\ref{Z-t-main-2}). We note that, unlike the numerical transfer matrix method, the numerical Grassmann integral approach only involves integer arithmetic, and
thus numerical floating-point errors are never an issue. Another difference is that the Grassmann integral approach can also be applied to 
graphs without any repeated structure. (See however \cite{bedjac} for a discussion of a generalized transfer matrix approach that also has these properties.)

\section*{Acknowledgements}

We acknowledge use of the C++ linear algebra library \textit{Armadillo} \cite{armadillo} for the numerical transfer matrix diagonalizations discussed in Sec. \ref{num}.

\appendix

\section{Proof of Eq. (\ref{Z-gf})}
\label{proof}

We start by defining the auxiliary quantity
\be
\Theta(\ell) = \left(\prod_{k=1}^{\ell} \frac{\partial }{\partial \log w_{\alpha_k}}\right){\cal Z}^q
\ee
where $\ell=1,2,\ldots,|E|$. Thus from Eq. (\ref{Zqcol2}) we have
\be
Z = \Theta(|E|)\big|_{w_{\alpha}=1}.
\label{Zfromtheta}
\ee
Explicit evaluation gives 
\begin{eqnarray}
\Theta(1) &=& {\cal Z}^q\, q\,{\cal G}^{(1)}_c(\alpha_1),\\
\Theta(2) &=& {\cal Z}^q \left[q^2\, {\cal G}^{(1)}_c(\alpha_1){\cal G}^{(1)}_c(\alpha_2) + q\,{\cal G}^{(2)}_c(\alpha_1,\alpha_2)\right], \\
\Theta(3) &=& {\cal Z}^q \Big[ q^3\, {\cal G}_c^{(1)}(\alpha_1){\cal G}_c^{(1)}(\alpha_2){\cal G}_c^{(1)}(\alpha_3)  + q^2 \big({\cal G}_c^{(2)}(\alpha_1,\alpha_2){\cal G}_c^{(1)}(\alpha_3) + {\cal G}_c^{(2)}(\alpha_1,\alpha_3){\cal G}_c^{(1)}(\alpha_2) \nonumber \\ &+& {\cal G}_c^{(2)}(\alpha_2,\alpha_3){\cal G}_c^{(1)}(\alpha_1)\big) + q\,{\cal G}_c^{(3)}(\alpha_1,\alpha_2,\alpha_3)\Big].
\end{eqnarray} 
From these expressions a pattern can be discerned, which suggests the following ansatz for $\Theta(\ell)$:
\be
\Theta(\ell) = {\cal Z}^q \sum_{P_{\ell}=1}^{B_{\ell}}q^{N_{P_{\ell}}}\prod_{r_{\ell}=1}^{N_{P_{\ell}}}{\cal G}_c^{(m_{P_{\ell}r_{\ell}})}(S_{P_{\ell}r_{\ell}}). 
\label{guess}
\ee
The sum is over all partitions of the edge set $\{\alpha_1,\alpha_2,\ldots,\alpha_{\ell}\}\equiv E_{\ell}$, which we label $P_{\ell}=1,2,\ldots,B_{\ell},$ where the Bell number $B_{\ell}$ is the total number of partitions.  Furthermore, $r_{\ell}=1,2,\ldots,N_{P_{\ell}}$ labels the subsets $S_{P_{\ell}r_{\ell}}$ of partition $P_{\ell}$, with $N_{P_{\ell}}$ the total number of subsets, $m_{P_{\ell}r_{\ell}}$ the number of edges in $S_{P_{\ell}r_{\ell}}$, and ${\cal G}^{(m_{P_{\ell}r_{\ell}})}_c(S_{P_{\ell}r_{\ell}})$ the connected dimer correlation function for this subset. 

We will prove (\ref{guess}) by induction. Assuming it holds for $\ell$ edges (the set $E_{\ell}$), we consider the case of $\ell+1$ edges (the set $E_{\ell+1}\equiv E_{\ell} \cup \{\alpha_{\ell+1}\}$). We have
\begin{eqnarray}
\lefteqn{\Theta(\ell+1) = \frac{\partial}{\partial \log w_{\alpha_{\ell+1}}}\Theta(\ell) = \frac{\partial}{\partial \log w_{\alpha_{\ell+1}}} \left[{\cal Z}^q \sum_{P_{\ell}=1}^{B_{\ell}}q^{N_{P_{\ell}}}\prod_{r_{\ell}=1}^{N_{P_{\ell}}}{\cal G}_c^{(m_{P_{\ell}r_{\ell}})}(S_{P_{\ell}r_{\ell}})\right]}  \nonumber \\ &=& 
{\cal Z}^q\, q\, {\cal G}_c^{(1)}(\alpha_{\ell+1}) \sum_{P_{\ell}=1}^{B_{\ell}}q^{N_{P_{\ell}}}\prod_{r_{\ell}=1}^{N_{P_{\ell}}}{\cal G}_c^{(m_{P_{\ell}r_{\ell}})}(S_{P_{\ell}r_{\ell}}) + 
{\cal Z}^q \sum_{P_{\ell}=1}^{B_{\ell}}q^{N_{P_{\ell}}} 
\frac{\partial }{\partial \log w_{\alpha_{\ell+1}}}\prod_{r_{\ell}=1}^{N_{P_{\ell}}} {\cal G}_c^{(m_{P_{\ell}r_{\ell}})}(S_{P_{\ell}r_{\ell}}) \nonumber \\ & = &    
{\cal Z}^q \sum_{P_{\ell}=1}^{B_{\ell}}q^{N_{P_{\ell}}}\Big[q\,{\cal G}_c^{(1)}(\alpha_{\ell+1})\prod_{r_{\ell}=1}^{N_{P_{\ell}}}{\cal G}_c^{(m_{P_{\ell}r_{\ell}})}(S_{P_{\ell}r_{\ell}}) 
\nonumber \\ &+& 
\sum_{t_{\ell}=1}^{N_{P_{\ell}}}{\cal G}^{(m_{P_{\ell}t_{\ell}}+1)}_c(S_{P_{\ell}t_{\ell}}\cup\{\alpha_{\ell+1}\})\prod_{\stackrel{r_{\ell}=1}{r_{\ell}\neq t_{\ell}}}^{N_{P_{\ell}}}{\cal G}_c^{(m_{P_{\ell}r_{\ell}})}(S_{P_{\ell}r_{\ell}})\Big].
\label{Thetares}
\end{eqnarray}
This expression is in fact of the form (\ref{guess}) with $\ell$ replaced by $\ell+1$. To see this, first note that the partitions of $E_{\ell+1}$ can be generated by
starting from those of $E_{\ell}$ and ``adding" the additional edge in all possible ways. For a given partition $P_{\ell}$ of $E_{\ell}$ we can either let the additional edge
be a subset by itself [this corresponds to letting $\partial/\partial \log w_{\alpha_{\ell+1}}$ act on ${\cal Z}^q$, which gives a factor ${\cal Z}^q\,q\,{\cal G}^{(1)}(\alpha_{\ell+1})$; 
this produces the first term in Eq. (\ref{Thetares})], or we can add it to any one of the $N_{P_{\ell}}$ subsets in that partition [this corresponds to letting $\partial/\partial \log w_{\alpha_{\ell+1}}$ act on the connected dimer correlation function for that subset, which gives the connected dimer correlation function for the union of the original subset and the new edge; this produces the second term in Eq. (\ref{Thetares})]. It follows that the total number of partitions $B_{\ell+1}$ of the set $E_{\ell+1}$ is given by
\be
B_{\ell+1} = \sum_{P_{\ell} =1}^{B_{\ell}}(1+N_{P_{\ell}}).
\label{partrel}
\ee
Eq. (\ref{partrel}) can be verified by invoking the so-called Stirling number of the second kind, $S(\ell,k)$, defined as the number of ways to partition a set of $\ell$ objects into $k$ (non-empty) subsets. From this definition it follows that the Bell number $B_{\ell}$ can be written $B_{\ell} = \sum_{k=0}^{\ell}S(\ell,k)$ and furthermore that 
$\sum_{P_{\ell}=1}^{B_{\ell}} N_{P_{\ell}} = \sum_{k=0}^{\ell} k S(\ell,k)$. Eq. (\ref{partrel}) can then be shown to follow from the recurrence relation
$S(\ell+1,k) = S(\ell,k-1)+ k S(\ell,k)$. 

Thus we have proved that if Eq. (\ref{guess}) holds for $\ell$, it also holds for $\ell+1$. Since Eq. (\ref{guess}) is true for $\ell=1$, the
induction proof of Eq. (\ref{guess}) is complete. Eq. (\ref{Z-gf}) then follows from Eq. (\ref{Zfromtheta}). 

\section{On our implementation of Creutz's algorithm}
\label{impl}

In this Appendix we make some remarks on our implementation of Creutz's algorithm for the numerical evaluation of the Grassmann integrals in Eq. (\ref{Z-t-main-2}) for the problem of a square lattice embedded on a torus discussed in
Sec. \ref{num} (these remarks are not likely to be very comprehensible unless the reader is already familiar with the discussion of the algorithm in Ref. \cite{creutz}).

The $x$ ($y$) direction is taken to be the transverse (longitudinal) direction in the algorithm. Memory requirements grow exponentially with $N_x$, the number of vertices in the transverse direction. 
Moreover, the growth rate for our problem is large due to (i) the large effective number of Grassmann variables per vertex, and (ii) the periodic boundary conditions in the longitudinal direction (which effectively doubles the growth rate compared to the case of open boundary conditions \cite{creutz}). Efficient use of memory is therefore important. We now describe some simple ways to reduce the memory needed.

Regarding (i): The effective number of Grassmann variables per vertex is naively 8, since there are 4 Grassmann variables living on each vertex, and 2 Grassmann variables living on each of the 2 edges that can be associated with a given vertex. 
The algorithm introduces one fermionic single-particle state for each Grassmann variable. Applying the algorithm to our problem it can however be seen that the fermions in the two single-particle states on a given edge are always annihilated together. It is then possible to modify the algorithm slightly so that for each edge one only needs to introduce a single fermionic single-particle state. This reduces the effective number of Grassmann variables per vertex to 6 (which is still large). 

Regarding (ii): The memory requirements can be considerably reduced by splitting the calculation of the Grassmann integral up into different parts. Let $v=1,\ldots,6N$ refer to a Grassmann variable ($N=N_x N_y$). The variables are labeled
according to the vertex they are associated with, with vertices 1 and $N$ in the bottom left and top right corners respectively. 
The Grassmann integral can be written \cite{creutz} $\langle 0|\psi_{6N}\rangle$ where $|0\rangle$ is the state with no fermions and $|\psi_v\rangle = Q_v|\psi_{v-1}\rangle$, where $Q_v$ is an operator and $|\psi_0\rangle = |F\rangle$, the completely
occupied state. Consider the state $|\psi_{6N_x}\rangle$ obtained after integrating out all $6N_x$ Grassmann variables associated with vertices in the bottom row. Each basis state with a nonzero coefficient in $|\psi_{6N_x}\rangle$ will have the following property: for each vertex in the \textit{top} row (which is, due to the periodic boundary conditions, coupled to the vertex in the bottom row directly ``above" it), one of the 4 fermionic states (each of which is associated with a different colour) will have occupation number 0, while the 3 other states will have occupation number 1. Thus one can split the terms in $|\psi_{6N_x}\rangle$ into distinct parts characterized by which of these states have occupation number 0, i.e. one writes 
$|\psi_{6N_x}\rangle = \sum_p |\psi_{6N_x,p}\rangle$ where $p$ labels the different parts. (The splitting can be done to various degrees, depending on one's needs: 
the coarsest splitting is into only 4 parts based on the states on just a single vertex in the top row, while the finest splitting would be into $4^{N_x}$ parts based on the states on all $N_x$ vertices.) Next one considers each part $p$ in turn, 
calculating $|\psi_{6N,p}\rangle = (\prod_{v=6N_x+1}^{6N}Q_v)|\psi_{6N_x,p}\rangle$ where the $Q_v$'s are applied in ascending order of $v$. The final result is obtained as $\langle 0|\psi_{6N}\rangle = \sum_p \langle 0|\psi_{6N,p}\rangle$. 
Doing this splitting has the advantage that the calculation of $|\psi_{6N,p}\rangle$ involves a considerably smaller maximum size of the hash table than would a direct calculation of $|\psi_{6N}\rangle$ (i.e. without doing this splitting). 
This is because the calculations of $|\psi_{6N,p}\rangle$ for different $p$'s have no states in common during those stages of the calculations when the hash tables reach their maximum sizes (which happens for specific intermediate values of $v$, 
not too close to the bottom or top rows). 

Each key-value pair in the hash table (``dictionary") consists of a many-particle fermionic basis state and its coefficient, both of which are integers (for the basis state, its binary representation 
gives the occupation numbers in each single-particle fermion state) whose size will typically exceed the computer's limit. Thus one needs a way to do arbitrary-precision integer arithmetic. We wrote the code in Python, which has built-in
support for both hash tables and arbitrary-precision integer arithmetic. The code was run on various computers with 2.4-3 GHz processors and 4-16 GB RAM. For $N_x=3$ the calculation of a Grassmann integral in Eq. (\ref{Z-t-main-2}) 
took a few seconds or less, while for $N_x=7,8$ it took about 10 days for the largest values of $N_y$ considered. 

\section{Unified expression for $W=\lim_{N\to\infty}Z^{1/N}$ for some lattices}
\label{unifying}

The problem of enumerating the $q$-edge-colourings for a lattice with degree (i.e. coordination number) $q$ has been solved exactly, in the thermodynamic limit, only 
for a relatively small number of lattices: the honeycomb lattice \cite{baxter-70}, the square lattice \cite{dc-nien-04} (see also \cite{jac-zj}), the 4-8 lattice \cite{jof10}, and the 3-12 lattice \cite{jof10}. 
These lattices have $q=3$ except the square lattice which has $q=4$. In addition the problem can
be solved trivially for a one-dimensional lattice ($q=2$). For a given lattice the exponential scaling of the number of $q$-colourings $Z$ is given by the parameter $W\equiv \lim_{N\to\infty}Z^{1/N}$, where $N$ is the number of lattice sites (vertices). 
In the following we demonstrate a connection between the results for $W$ for the honeycomb, square, 3-12, and the one-dimensional lattice. 

For the honeycomb (hc) lattice, it was found that \cite{baxter-70,baxter-86}
\be
W_{\rm{hc}}^2 = \prod_{j=1}^{\infty}\frac{(3j-1)^2}{3j(3j-2)} = \frac{3\Gamma^3(1/3)}{4\pi^2},
\ee
while for the square (sq) lattice, it was found that \cite{dc-nien-04}
\be
W_{\rm{sq}} = \prod_{j=1}^{\infty} \frac{(4j-1)(2j-1)}{2j(4j-3)}=\frac{\Gamma^2(1/4)}{\sqrt{2}\,\pi^{3/2}}.
\ee
For the 3-12 lattice, $W_{\rm{3-12}}=W^{1/3}_{\rm{hc}}$ \cite{jof10}. Finally, as the one-dimensional (1D) lattice has $Z=2$ for all (even) $N$, it follows that $W_{\rm{1D}}=1$. 

The results for the parameter $W$ for these four lattices can be written in a unified form in the following equivalent ways:
\begin{eqnarray}
W^p &=& \prod_{j=1}^{\infty}\frac{(qj-1)(qj-q+2)}{qj(qj-q+1)} = \frac{\Gamma(1/q)}{\Gamma(1-1/q)\Gamma(2/q)} \nonumber \\ &=& 
\frac{\sin(\pi/q)}{\pi}B(1/q,1/q) = \;_2 F_1\left(1/q,1-2/q;1;1\right),
\label{unif}
\end{eqnarray}
where $\Gamma(x)$ is the gamma function, $B(x,y)$ is the beta function, and $_2 F_1(a,b;c;z)$ is Gauss's hypergeometric function.\footnote{In order to convert between the different expressions in (\ref{unif}) it is convenient to insert the factor $\Gamma(1)=1$ in the numerator in the expression involving gamma functions.} Each lattice is associated with 
a pair of numbers $(q,p)$ where $q$ is the degree/coordination number as before and the value of $p$ coincides with the number of sites per unit cell of the lattice. 
Thus $(q,p)_{\rm{hc}}=(3,2)$, $(q,p)_{\rm{sq}}=(4,1)$, $(q,p)_{\rm{3-12}}=(3,6)$, and $(q,p)_{\rm{1D}}=(2,1)$.  

On the other hand, for the 4-8 lattice, it was found that \cite{jof10}
\be
W_{\rm{4-8}}=x^{-1/4}\prod_{j=1}^{\infty}\left(\frac{1-x^{4j-1}}{1-x^{4j+1}}\right)^{1/2}
\label{W48}
\ee
where $x=2-\varphi$ with $\varphi=(1+\sqrt{5})/2$ being the golden ratio. More recently this value for $W_{\rm{4-8}}$ was also verified numerically \cite{num48}. 
However, using $(q,p)_{\rm{4-8}}=(3,4)$ in (\ref{unif}) fails to reproduce this value. Of course, one might question whether the interpretation of $p$ as the number of sites per unit cell is really the correct one. 
But keeping $q=3$, numerical agreement between (\ref{unif}) and (\ref{W48}) would require one to take $p\approx 1.759$, a value that it is difficult to make sense of, also in light of the ``natural" values for $p$ required for the other lattices. 
Thus we conclude that (\ref{unif}) is not valid for the 4-8 lattice. 

A deeper understanding of (\ref{unif}) will have to await a general derivation of it for a class of lattices that should include the honeycomb, square, 3-12, and one-dimensional lattice,
and possibly other lattices as well that have not been identified yet.


 \end{document}